\documentclass[11pt,twoside]{article}

\usepackage{asp2006}
\usepackage{epsf}
\usepackage{lscape}
\usepackage{graphicx}

\begin{document}
\title{Inverse-Compton Emission from the Lobes of 3C353}  
\author{J.L. Goodger, M.J. Hardcastle, and J.H. Croston}  
\affil{University of Hertfordshire, Science and Technology Research Institute, Hatfield, Herts AL10 9AB, UK}   

\begin{abstract} 

Analysis of the radio synchrotron and X-ray inverse-Compton emission
from radio-loud active galaxies allows us to determine their particle
acceleration processes and electron energy spectra. Previous studies
have provided new constraints on the total energy budget and particle
content of powerful radio galaxies and quasars; however, in most cases
the sources are too faint in the X-ray to obtain spatial information.
We present archival and new multi-frequency radio observations from
the VLA and GMRT, and {\it XMM-Newton} observations of the bright FRII
radio galaxy 3C353 which lies on the edge of the X-ray-luminous
cluster Zw 1718.1-0108. The X-ray observations detect both the
inverse-Compton emission from the radio galaxy lobes and thermal
emission from the hot phase of the intracluster medium.  We discuss
the properties of the particle energy spectrum as a function of
position in the lobe, as well as the properties of the merging
cluster.

\end{abstract}

\section{Introduction} 

By studying radio galaxies in the radio and X-ray, we can determine
the densities of the electrons emitting via synchrotron radiation in
the radio and inverse-Compton in the X-ray.  The radio emission alone
cannot separate the effects of electron density and magnetic field
strength; however using X-ray data, the electron energy density at low
energies can be measured directly.  The first detections of radio
lobes in the X-ray were identified as inverse-Compton scattering of
the CMB \citep[e.g.][]{feigelson95jlg, hardcastle02jlg} or of nuclear
photons \citep[e.g.][]{brunetti97jlg}.  More recent studies of the
integrated X-ray properties of radio lobes have now been carried out
for large samples \citep{croston05jlg, kataoka05jlg}, showing that the
CMB is the dominant photon population in most cases and providing
measurements of the total energy in radio source and hence the work
done on the environment.  Inverse-Compton emission can now be used to
investigate the spatial distribution of electrons and magnetic field
in FRII radio galaxies \citep[e.g.][]{hardcastle05jlg}.  Here we use
spatially resolved X-ray emission to carry out such a study of 3C353.

\begin{figure}[!t]
\plotfiddle{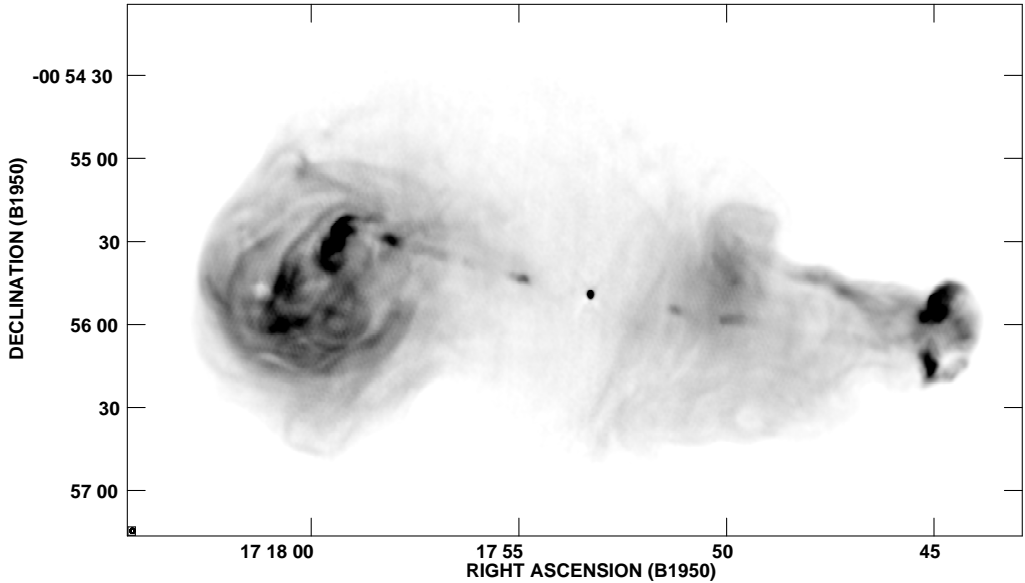}{70mm}{270}{55}{55}{-220}{275}
\caption{1.4 GHz total intensity image of 3C353 at 1.8 arcsec resolution}
\label{fig:radio-jlg}
\end{figure}

\begin{figure}[!t]
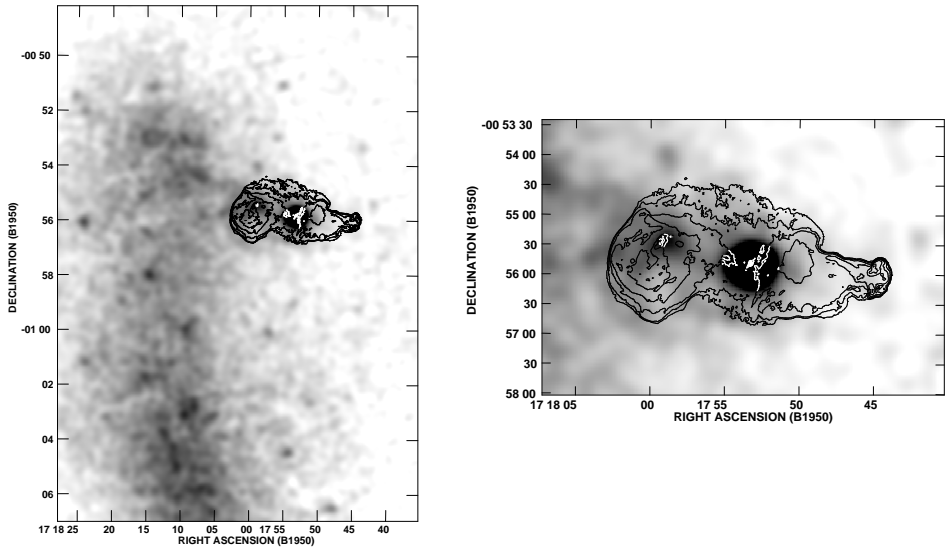

\plotfiddle{3C353-xrbw-jlg.ps}{30mm}{0}{30}{30}{-200}{-120}
\plotfiddle{3C353-xrbwrgp-jlg.ps}{30mm}{0}{35}{35}{-30}{-40}
\caption{Radio contours of 3C353 at 1.4 GHz with a Gaussian smoothed 0.3-7.0 keV {\it XMM-Newton} image (MOS 1, MOS 2 + pn) of the X-ray emission from the radio lobes and the cluster Zw 1718.1-0108 (left), zoomed in to show the inverse-Compton emission from the lobes (right).} 
\label{fig:xray-jlg}
\end{figure}

3C353 is an FRII, radio-loud active galaxy associated with the cluster
Zw 1718.1-0108 at redshift $z=0.030$.  The radio source resides in a
giant elliptical galaxy on the edge of the cluster. The cluster is
bright in the X-ray with a luminous point source coinciding with the
radio galaxy's core \citep{iwasawa00jlg}.  As one of the brightest
object in the low-frequency sky, 3C353 is an ideal target for an
inverse-Compton study.  We observed it at 330 MHz with the VLA and at
614 MHz at the GMRT in addition to using VLA archival data at 1.4 GHz,
5 GHz and 8.4 GHz, all of which were calibrated using standard
methods.  Figure~\ref{fig:radio-jlg} shows an example of the image
quality achieved in the radio.  We also observed 3C353 in the X-ray
with {\it XMM-Newton} (49408 s livetime for MOS 1 and MOS 2 and 31704
s for pn), so the broad-spectrum electron energy distribution is well
sampled.  We also resolve the X-ray thermal emission from the cluster,
shown in Figure~\ref{fig:xray-jlg}, which fills the field of view of
{\it XMM-Newton}.  Our radio data extends to lower frequencies than
have been studied previously in the published literature
\citep*{swain98jlg} whilst our X-ray data has greatly improved
sensitivity and resolution compared to the {\it ASCA} data of
\citet{iwasawa00jlg}.

\section{Origin of the X-ray Emission}

By considering the X-ray emission from regions encompassing the east
and west lobes but excluding the hotspots and the core, we extracted
spectra from the X-ray data set which we fitted with both thermal
emission models and power-laws. The power-law model gave better fits
than the thermal model, with $\chi^{2} = 45.9$ for 44 d.o.f for the
east lobe and $\chi^{2} = 23.7$ for 27 d.o.f for the west lobe.  The
photon indices were $2.0 \pm 0.2$ and $1.8 \pm 0.4$ respectively,
consistent with an inverse-Compton model.  We therefore conclude that
the X-ray emission from the lobes is non-thermal in origin.

\begin{figure}[!ht]
\plotfiddle{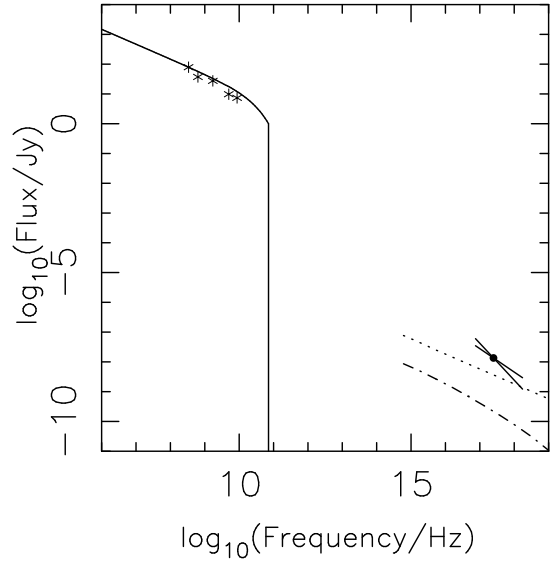}{70mm}{0}{50}{50}{-150}{-30}
\caption{Broad-band spectrum for the east lobe of 3C353. A synchrotron emission model (solid line) is fitted to the radio data (asterisks). The dotted line shows the inverse-Compton X-ray flux prediction and the dash-dotted line shows the predicted synchrotron self-Compton emission, both with the equipartition magnetic field strength.}
\label{fig:synch-jlg}
\end{figure}

Using the multi-frequency radio data, we modelled the synchrotron
emission and predicted the form of the inverse-Compton emission at
equipartition.  The broad-band spectrum for the east lobe is shown in
Figure~\ref{fig:synch-jlg}.  The equipartition inverse-Compton X-ray
flux prediction for both lobes is less than the observed X-ray flux by
a factor of ~2.  This corresponds to a variation in the magnetic field
strength, $B_{obs}/B_{eq} = 0.4$.  This difference lies within the
range of this factor determined by \citet{croston05jlg} using a sample
of 33 FRII galaxies and quasars and indicates that the lobes of 3C353
are electron dominated.  This factor would be reduced if the lobes
were not in the plane of the sky.

\section{Lobe Properties as a Function of Position}

Constructing a spectral index map between the 1.4 GHz and 330 MHz
images revealed a variation of $\alpha \sim 0.1$ across the east lobe.
Within the radio luminous region of the east lobe, excluding the
hotspots, the spectral index is roughly constant, $\sim 0.66$, despite
the filamentary structure seen in Figure~\ref{fig:radio-jlg}.  We note
that the radio lobes do not appear to be entirely separate, and thus
consider the region in between the radio luminous lobes, north and
south of the core to be an inter-lobe region, which cannot be
unambiguously associated with either lobe.  This inter-lobe region has
a relatively steep radio spectrum whereas the region around the east
hotspot exhibits a flatter spectrum.

We also observe a variation in the X-ray/radio ratio, with the steep
spectrum lobe region having a ratio four times that of the flat
spectrum region near the hotspot. If the magnetic field strength and
the number densities of both the inverse-Compton and synchrotron
emitting electrons were constant across the lobe, the X-ray/radio
ratio would be constant. As this is not what we observe, one or both
of these quantities must depend on position in the lobe. We therefore
examined the effect of varying either the magnetic field strength or
the number densities as a function of position.

\subsection{Electron Spectrum Variation}

If we consider the magnetic field to be constant at the observed value
(see Section 2) and apply an inverse-Compton model to the lobes, we
find that the emission at 330 MHz traces electrons with $\gamma = 4500$
whilst the inverse-Compton emission traces electrons at $\gamma =
1000$.  As the critical frequency for synchrotron emission goes as
$\gamma^{2}$, we find that a variation in the X-ray/radio ratio of a
factor of 4 would require a variation in the synchrotron spectral
index between 10 MHz and 330 MHz of $\sim 0.46$ as opposed to the
observed $\Delta \alpha$ at higher frequencies of $\sim 0.1$.  Even if
the equipartition magnetic field strength is assumed so that the 330
MHz emission traces electrons of $\gamma \sim 3000$, the observed
spectral index variation requires a factor $\sim 1.2$ variation in the
X-ray/radio ratio which is much lower than observed.  We cannot
therefore explain the observed variations with varying electron
densities alone.

\subsection{Magnetic Field Variation}

If we assume constant electron densities for both the synchrotron and
inverse-Compton emission electrons and also assume that the magnetic field
strength does not vary along the line of sight, we find that the
observed variation in the X-ray/radio ratio then requires a variation
in the magnetic field of at most a factor 2.5.

For a given electron population, the critical frequency goes as $B$,
so that the spectral indices observed at low frequencies, between 1.4
GHz and 330 MHz, for regions of high X-ray/radio ratio correspond to
the spectral indices between 3.5 GHz and 825 MHz for low X-ray/radio
ratio regions.  We consider the spectral indices between 5 GHz and 1.4
GHz to limit the spectral indices of the low X-ray/radio ratio regions
and find that they exceed the upper limit predicted by the observed
X-ray/radio ratio.

We thus cannot rule out the possibilty that a varying magnetic field
alone could be responsible for the observed variations in the
X-ray/radio ratio and the spectral indices; however, we conclude that
a varying electron spectrum cannot account for these differences
without the addition of a magnetic field variation.  Similar
conclusions were reached by \citet{hardcastle05jlg} in a study of
Pictor A.  Analysis of the detailed radio spectral properties will
help us to see whether a model in which the electron spectrum is
constant throughout the lobes is viable.

\section{Characterisation of the Cluster}

Figure~\ref{fig:xray-jlg} also shows the thermal emission from the
cluster.  Thermal fitting to the X-ray data sets for the north and
south subcluster regions have revealed a difference in mean
temperature, with the northern subcluster having $kT = 2.8 \pm 0.3$
keV and the southern $kT = 3.8 \pm 0.3$ keV.  There appears to be no
noticable increase in temperature at the boundary, suggesting
that the north and south subclusters are not interacting violently.

\citet{iwasawa00jlg} included optical observations of the cluster
region centred on 3C353, taken using the University of Hawaii 2.2m
telescope, in their analysis of Zw 1718.1$-$0108. They identified
three additional large galaxies, none of which resides in the centre
of the northern sub-cluster.  Deeper optical observations would help
to establish whether there are any galaxies associated with the
northern X-ray subcluster.  

Further analysis of the cluster X-ray emission and lobe-cluster
interaction is ongoing.  We will use surface brightness profiles to
measure the densities of the clusters which, when combined with the
temperature profiles, will allow us to determine the physical
conditions in the cluster and the dynamics of the lobe-cluster
interaction (Goodger et al., in prep.).

\acknowledgements 

JLG thanks the STFC for a research studentship. MJH acknowledges
generous financial support from the Royal Society.


\end{document}